\documentstyle[aps,multicol,pre,epsfig]{revtex}

\newcommand{\beq}{\begin{equation}}
\newcommand{\eeq}{\end{equation}}
\def\FigSize{8.75cm}

\begin{document}

\begin{title}
{\bf Stability of dark solitons in a Bose-Einstein condensate
trapped in an optical lattice}
\end{title}

\author{P.G.\ Kevrekidis$^{1}$, R.\ Carretero-Gonz\'alez$^2$,
G.\ Theocharis$^3$, D.J.\ Frantzeskakis$^{3}$, and B.A.\ Malomed$^4$}
\address{$^{1}$ Department of Mathematics and Statistics,
University of Massachusetts, Amherst MA 01003-4515, USA \\
$^2$ Nonlinear Dynamical Systems Group, Department of Mathematics \& Statistics,\\
San Diego State University, San Diego CA, 92182-7720, http://nlds.sdsu.edu/,%
\\
$^{3}$ Department of Physics, University of Athens, Panepistimiopolis,
Zografos, Athens 15784, Greece \\
$^{4}$ Department of Interdisciplinary Studies, Faculty of
Engineering, Tel Aviv University, Tel Aviv 69978, Israel
\\[1.0ex] {\sl Phys.\ Rev.\ A {\bf 68} 035602 (2003)}
}
\maketitle

\begin{abstract}
We investigate the stability of dark solitons (DSs) in an effectively
one-dimensional Bose-Einstein condensate
in the presence of the magnetic parabolic trap and an optical lattice (OL).
The analysis is based on both the full Gross-Pitaevskii equation
and its tight-binding approximation counterpart (discrete nonlinear
Schr{\"o}dinger equation).
We find that DSs are subject to weak instabilities with
an onset of instability mainly governed by
the period and amplitude of the OL.
The instability, if present, sets in at large times
and it is characterized by quasi-periodic oscillations of the DS about the
minimum of the parabolic trap.
\end{abstract}

\begin{multicols}{2}

\vspace{2mm}The experimental creation and great advancement in the
theoretical understanding of Bose-Einstein condensates (BECs) \cite{review}
have stimulated a lot of interest in nonlinear matter waves, including
dark \cite {dark} and bright \cite{bright} solitons.
The dynamics of dark solitons (DSs) in the presence of the external
magnetic trap has been extensively studied \cite{motion},
including thermal \cite{dis} and dynamical \cite{thsnbec} instabilities.
More recently, apart from the rectilinear DSs,
ring-shaped counterparts were predicted in BECs \cite{george}.

The study of nonlinear excitations is particularly relevant for BECs trapped
in optical lattices (OLs) generated by interference patterns from laser beams
illuminating the condensate \cite{g2,g3,g4,g5,g6,g7}.
The controllable character of the OL allows for
the observation of numerous phenomena, such as Bloch oscillations \cite
{g4,g8} and Zener tunneling \cite{g2} (in the presence of an additional
linear external potential), or classical \cite{Catall} and quantum \cite{g7}
superfluid-insulator transitions.

Apart from the mean-field description via the Gross-Pitaevskii (GP)
equation \cite{g2,g3,g4,g5,g6,g7},
a BEC trapped in a strong OL may be described, in the tight-binding limit,
by the discrete nonlinear Schr{\"{o}}dinger (DNLS) equation \cite{tromb}.
This approximation is not always accurate, but its applicability can be
systematically examined \cite{wannier}. In cases where such a reduction is
possible (e.g., when the chemical potential is much lower than the height of
the potential barriers induced by the OL), the DNLS model is particularly
relevant and has been successfully applied in many instances (for a
recent review on DNLS see, e.g., \cite{intn} and references therein).

In this paper, we study DSs in repulsive BECs (i.e.,
positive-scattering-length collisions) in the presence of OLs.
We use both the continuous-GP and DNLS equations. In particular, we assume a
cigar-shaped BEC, which can be described by the following normalized
quasi-one-dimensional GP equation \cite{review,rupr,GPE1d},
\begin{equation}
iu_{t}=-u_{xx}+|u|^{2}u+\left[ kx^{2}+V_{0}\sin ^{2}\left( 2\pi x/\lambda
\right) \right] u\,.  \label{gpe}
\end{equation}
Here, $u(x,t)$ is the mean-field wave function, while the terms in the
square brackets represent the external magnetic trap and the OL potential,
respectively, with the strengths $k$ and $V_{0}$, while $\lambda $ is the
wavelength of the interference pattern created by the laser beams.

To study the dynamics of a DS in the framework of Eq.\ (\ref{gpe}), we
consider an initial condition similar to an ansatz proposed for the
description of DSs in BECs in Ref. \cite{Carlo}
\begin{equation}
u_{0}(x)=u_{{\rm TF}}(x)\,\tanh (x-x_{0}),  \label{initialDS}
\end{equation}
where $u_{{\rm TF}}=\sqrt{{\rm max}(0,\mu -kx^{2})}$ is the Thomas-Fermi
(TF) expression for the background wave-function distribution \cite{review}
($\mu $ is the chemical potential) and $x_{0}$ is the initial location of the
DS's center. In most cases, we set $x_{0}=0$, i.e., the dark soliton is
placed at the bottom of the magnetic trap.

In the tight-binding limit, Eq.\ (\ref{gpe}) reduces to the following DNLS
equation \cite{tromb},
\begin{equation}
i\dot{u}_{n}=-C\left( u_{n+1}+u_{n-1}-2u_{n}\right)
+|u_{n}|^{2}u_{n}+kn^{2}u_{n}.  \label{dnls}
\end{equation}
where the dot denotes time derivative, $n$ is the lattice-site index, and $C$
is the so-called coupling constant (see e.g., Refs. \cite{tromb,wannier} for
exact expressions and relevant estimates). An initial condition for a DS in
the case of Eq.\ (\ref{dnls}) can be given by a straightforward
discretization of the
continuum ansatz (\ref{initialDS}). Typically, simulations were run for a
lattice with $200$ sites, and free boundary conditions were used for both
the continuum and discrete models. In fact, it has been verified that the
results are insensitive to the choice of boundary conditions. Note that DSs in
discrete lattices (in the absence of a parabolic trap)
were already studied in the framework of the DNLS equation
\cite{KKC}, revealing that they are subject to oscillatory
instabilities \cite{JK}.

Stationary solutions to Eq.\ (\ref{dnls}) are sought for in the form
$u_{n}=\exp (-i\mu t)v_{n}$, where $\mu $ is the chemical potential
\cite{review}, which leads to the steady-state equation,
\begin{equation}
-C\left( v_{n+1}+v_{n-1}-2v_{n}\right) +(|v_{n}|^{2}+kn^{2}-\mu )v_{n}=0,
\label{rdnls}
\end{equation}
for a discrete function $v_{n}$. Once a solution of Eq.\ (\ref{rdnls})
is found, linear stability analysis is performed by looking for
perturbed solutions of the form
\begin{equation}
u_{n}=e^{-i\mu t}\, [ v_{n}+\epsilon a_{n}e^{i\omega t}+\epsilon
b_{n} e^{-i\omega ^{\ast }t}],  \label{l}
\end{equation}
where the asterisk denotes complex conjugation. The ensuing eigenvalue
problem, for the eigenfrequency-eigenfunction pair
$\{\omega,\{a_{n},b_{n}^{*}\}\}$), is then solved numerically with
$\omega =\omega _{r}+i\omega _{i}$ (where the
subscripts denote the real and imaginary parts).
In what follows, since Eqs.\ (\ref{gpe}) and (\ref{dnls}) admit
an additional rescaling,
the chemical potential has been fixed at $\mu =1$.

\begin{figure}[tbp]
\epsfxsize=\FigSize
\centerline{\epsffile{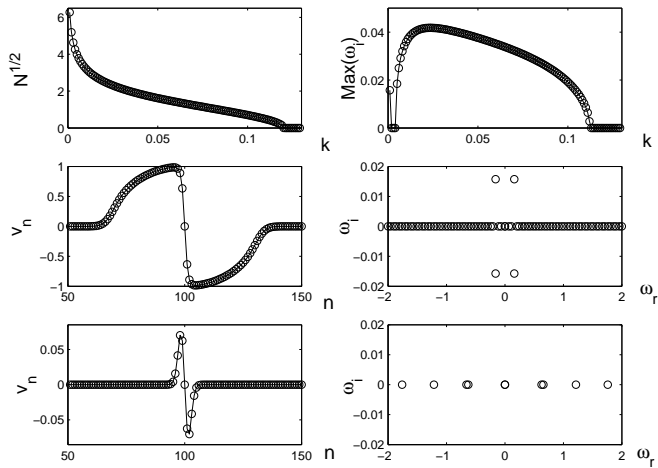}}
\caption{
The left and right top panels show, respectively, the (square root of the)
number of atoms in the condensate $N$ and the largest
unstable eigenfrequency $\omega _{i}$ for the discrete dark soliton as a
function of $k$ for $C=1$. The profile of the (exact)
solution is shown, for $k=0.001$ and
$k=0.119$ (just prior to the termination of the branch), in the left
middle and bottom panels, respectively. The right middle and bottom panels
show the spectral plane $(\omega _{r},\omega _{i})$ of the corresponding
eigenfrequencies, the subscripts standing for the real and imaginary parts.}
\label{dsfig1}
\end{figure}

We first consider the DNLS model with a fixed coupling constant $C=1$
while the strength $k$ of the parabolic trap is
varied. In this case, it is found that the assumed configuration, in the
form of a DS on top of the TF background, exists only up to a critical value
of $k_{{\rm cr}}\approx 0.12$, as is seen from the top left panel of Fig.\ \ref{dsfig1},
which complies with the well-known fact that the DS cannot have a width
essentially smaller than the healing length \cite{review}, and thus cannot
exist in a very narrow trap.

This solution family is quantified by the top panels of
Fig.\ \ref{dsfig1}. The norm
$N \equiv ||v_{n}||_{2}^{2}=\sum_{n}v_{n}^{2}$
(number of atoms $N$) as a function of
$k$ is depicted in the top left panel.
The dependence of the largest instability growth rate $\omega _{i}$ on $k$,
which is shown, for the same discrete-DS solution, in the top right panel of
Fig.\ \ref{dsfig1}, reveals a narrow stability window close to $k=0$ (for
$0.002\leq k\leq 0.004$). At larger values of $k$, the DS is always subject
to an oscillatory instability similar to that found for regular dark
optical solitons \cite{JK}. Nevertheless, this instability is characterized
by a {\em small} growth rate, whose maximum value is $\omega_{i}\approx 0.04$
(at $k\approx 0.024$).

An example of the weakly unstable DS for $k=0.001$ can be seen in the middle
panels of Fig.\ \ref{dsfig1}, and an example close to the termination of the
branch (at $k=0.119$) is shown in the bottom panels. Note that the
instability is manifested by the presence of a nonzero imaginary part of the
eigenfrequencies.

\begin{figure}[tbp]
\epsfxsize=\FigSize
\centerline{\epsffile{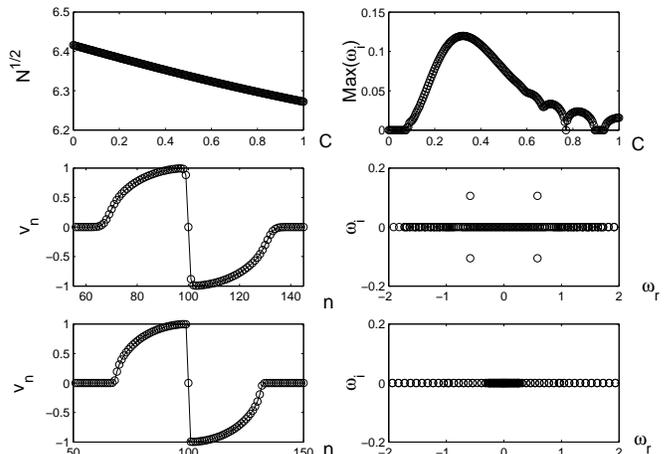}}
\caption{
The top panels in Fig.\ \ref{dsfig2} are equivalent to those in Fig.\ \ref
{dsfig1}, but now the strength of the trapping potential is fixed,
$k=0.001$, while the coupling constant $C$ is varied.
The middle and bottom panels
show, respectively, examples of the spatial profiles and eigenfrequency
spectral planes of unstable (for $C=0.25$) and stable (for $C=0.01$) DSs.}
\label{dsfig2}
\end{figure}

Next, we consider the case where the strength of the parabolic trap is kept
fixed, $k=0.001$, while the coupling constant $C$ is varied in the interval
$0\leq C\leq 1$. In this case, the DS configuration on top of the TF
background has been obtained
for every value of $C$, see Fig.\ \ref{dsfig2}, down to $C=0$
corresponding to the so-called anti-continuum (AC) limit.
The norm of the solution (top left panel of Fig.\ \ref{dsfig2}) is a slowly
(almost linearly) decreasing function of $C$, which can be easily
understood. Indeed, since the effective size of the TF distribution is kept
constant (the trap strength $k$ is fixed), the DS placed at the center of
the condensate expands to a larger number of lattice sites as $C$ increases,
hence a bigger internal part of the BEC cloud is effectively devoid of
atoms. 
The dependence of the largest unstable eigenfrequency $\omega _{i}$ on $C$,
shown in the top right panel of Fig.\ \ref{dsfig2}, reveals that there exist
a critical value, $C_{{\rm cr}}\approx 0.076$, at which a Hamiltonian-Hopf
bifurcation occurs, as two pairs of eigenvalues with opposite {\it Krein
signatures} \cite{krein} collide and bifurcate into a complex quartet.

Notice the proximity of the critical value observed herein and the one found
in Ref. \cite{JK} for regular optical DSs (without the parabolic trap).
This observation suggests that
the presence of the parabolic trap does not significantly
affect the threshold of the oscillatory instability.

It is noteworthy that, at larger values of $C$ (see the top right panel of
Fig.\ \ref{dsfig2}), there are windows on the $C$ axis (such as e.g.,
$0.895\leq C\leq 0.935$ for the 200-site lattice) where the DS is stable.
Examples of unstable (for $C=0.25$) and stable DSs (for
$C=0.01$), as well as the respective spectral planes
$(\omega _{r},\omega _{i})$ of
the eigenfrequencies, are shown in the middle and bottom panels of Fig.\ \ref
{dsfig2}. As noted in Ref. \cite{JK}, the stability windows result from
the finiteness of the lattice.

As the DNLS of Eq.\ (\ref{dnls}) is only a tight-binding approximation of the
full continuum model, i.e., the GP equation (\ref{gpe}), it is necessary to
investigate whether and how the weak instability of the DSs,
demonstrated above in the discrete limiting case, manifests itself in the
full GP equation. To this end, we first consider the case where the OL
potential is fixed (an analog of the case where $C$ is fixed in the
discrete model), with strength $V_{0}=1.5$ and wavelength $\lambda =5$.

\begin{figure}[tbp]
\epsfxsize=\FigSize
\centerline{\epsffile{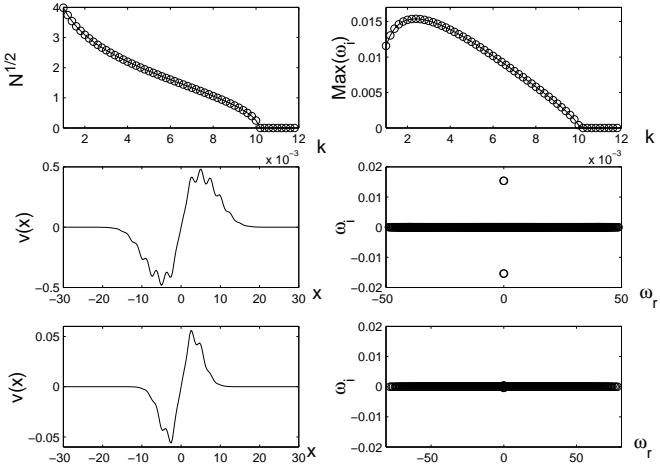}}
\caption{
The top panels show the number of atoms
rescaled (for convenience) by a factor of $10^{-1/4}$ (left) and the
maximum of the instability growth rate $\omega _{i}$ (right) for the dark
soliton versus the parabolic trap strength $k$, as found from the
continuum Gross-Pitaevskii equation (\ref{gpe}) with $V_{0}=1.5$ and
$\lambda =5$. The middle and bottom panels display the spatial profiles
for the dark solitons with $k=0.0024$ and $k=0.01$ (left) and the
respective stability spectral planes (right). [$k=0.01$ being very close to the
termination point of the branch.]}
\label{dsfig3}
\end{figure}

Similar to what we find for the discrete case, the DS on the TF
background exists, in the continuum GP equation, only up to a
critical value of the trap strength, $k_{{\rm cr}}\approx 0.01$. This is
demonstrated in the top left panel of Fig.\ \ref{dsfig3}, where
the number of atoms, which is now defined as
$
N=||v||_{2}^{2}\equiv\int_{-\infty }^{+\infty }\left| v(x)\right|^{2}dx\,,
$
is shown versus $k$. Furthermore, the eigenvalue computation shows
that this family of continuum DSs is always unstable (although in other
cases DSs in the continuum model may be stable, see below). However the
instability is {\em extremely weak}, as its largest growth rate
${\rm max}(\omega _{i})\approx 0.015$ (at $k\approx 0.024$).
The shape of typical continuum DSs are shown in the middle and bottom panels
of Fig.\ \ref{dsfig3} together with the corresponding spectral planes.

\begin{figure}[tbp]
\epsfxsize=\FigSize
\centerline{\epsffile{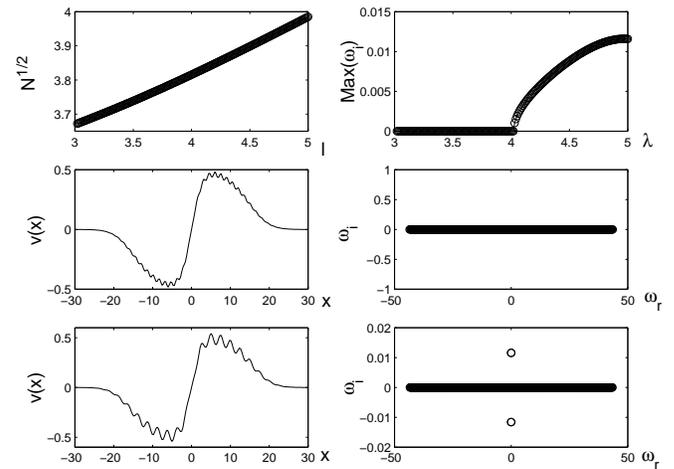}}
\caption{
The top panels show the number of atoms in the continuum dark soliton
(rescaled by the factor $10^{-1/4}$; left) and the largest instability
growth rate $\omega _{i}$ (right) versus the OL wavelength
$\lambda $ for the GP equation with
$V_{0}=1.5$ and $k=0.001$. The middle and bottom panels show the spatial
profile of the dark solitons for $\lambda =3$ and $\lambda =5$ (left) and their
respective spectral planes (right).}
\label{dsfig4}
\end{figure}

Next, we consider the case when the strength of the magnetic trap, in the
continuum GP equation, is fixed at $k=0.001$, while the wavelength $\lambda$
of the OL is varied (the strength of the OL is again
$V_{0}=1.5$). In this case, the DS on the TF background exists
for every value of $\lambda$ in an interval $3\leq \lambda \leq 5$,
which is chosen to display the present case. In particular, the top left
panel in Fig.\ \ref{dsfig4} shows the number of atoms in the DS as a function
of $\lambda $. Furthermore, the dependence of the largest instability growth
rate $\omega _{i}$ on $\lambda$ (top right panel in Fig.\ \ref
{dsfig4}) reveals that a Hamiltonian-Hopf bifurcation occurs at a critical
point, $\lambda _{{\rm cr}}^{(1)}\approx 4$, so that the DSs are
stable for $\lambda <\lambda _{{\rm cr}}^{(1)}$ and unstable otherwise.
Examples of stable and unstable DSs for $\lambda =3$ and $\lambda =5$ are
shown in the middle and bottom panels in Fig.\ \ref{dsfig4}, together with their
respective spectral planes.
We have verified that for $\lambda <3$ the DS retains
its stability, while for $\lambda >5$ it remains unstable up to $\lambda
<\lambda _{{\rm cr}}^{(2)}\approx 5.45$ and restabilizes for larger values
of $\lambda$. After running simulations at many other parameter values,
we have concluded that the OL periodicity $\lambda$ is a crucial factor
in determining the stability of the DS in the continuum GP equation, while
the strength of the parabolic trap $k$ plays a lesser role.

Finally, an additional factor of relevance in determining stability
is the amplitude $V_0$ of the optical lattice. We have found (data not
shown) that variation of $V_0$ results in the presence of windows of
stability for a fixed value of $\lambda$. For instance for $\lambda=5$,
the soliton is stable in the absence of the optical lattice (i.e., for
$V_0=0$) as well as e.g., in the intervals $V_0 \in [0.04,0.13]$ or
$V_0 \in [1.17,1.38]$, while it is unstable for intermediate values.

\begin{figure}[tbp]
\epsfxsize=\FigSize
\centerline{\epsffile{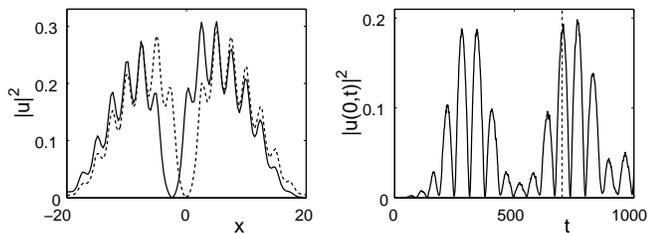}}
\caption{
Left: The profile of the dark soliton at $t=0$ (dashed line) and $t=700$
(solid line). Right: Time evolution of the density, $\left|
u(x,t)\right| ^{2}$, at the central point, $x=0$. $V_0=1.5$, $\lambda=5$ and
$k=0.001$.}
\label{dsfig5}
\end{figure}

In the case when solitons are unstable, it is desirable to directly simulate
the full nonlinear equation, in order to observe the evolution of the
soliton under the effect of the instability.
We find that the instability sets in at large times, on the order of
$O(100)$, and manifests itself as follows: the DS, which is
initially placed at the bottom of the composite potential (magnetic trap and
OL), performs small-amplitude oscillations inside the respective small well
of the OL potential. Then, as time passes, the soliton starts to emit small
amounts of radiation, and, as a result,
it gains enough kinetic energy to perform larger-amplitude
quasi-periodic oscillations around the center of the condensate. This
behavior is demonstrated in the left panel of Fig.\ \ref{dsfig5}, where two
snapshots of the density profile of the DS are shown at $t=0$ and $t=700$
for the unstable DS of the bottom panel of
Fig.\ \ref{dsfig4}. To trigger the onset of instability, a uniformly
distributed random perturbation of amplitude $0.01$ is added to the
initial configuration. It can be clearly observed that, in the final
state, the DS has shifted its position from
the center of the condensate ($x=0$). Quasi-periodic oscillations of the
density at the point $x=0$, induced by the instability, are shown in the
right panel of Fig.\ \ref{dsfig5}.

In conclusion, we investigate the existence and stability of
stationary dark solitons in repulsive BECs, trapped in an optical
lattice (OL). The consideration is performed in the framework of both the
continuous Gross-Pitaevskii (GP) equation and its discrete tight-binding
dynamical-lattice counterpart, taking into regard the effect of the external
magnetic trap and OL. We find that in the discrete model the dark solitons
are, generally, subject to a {\em weak} oscillatory instability.
In the full GP equation, dark solitons may be stable and the stability is
chiefly determined by the period and amplitude of the OL. If the oscillatory
instability is present, it sets in at large times, which attests to the
robustness of dark solitons. The instability eventually manifests itself as
a shift of the dark soliton from the center of the condensate, which is
accompanied by quasi-periodic oscillations.
This work can be naturally extended for  two-dimensional vortices. First
steps in that direction were made in Refs. \cite{we} and \cite{bsm}.

This work was supported by a UMass FRG and NSF-DMS-0204585 (PGK), the
Special Research Account of the University of Athens (GT, DJF), the
Binational (US-Israel) Science Foundation (No.\ 1999459),
the San Diego State University Foundation (RCG), and a
Windows-on-Science grant from the European Office of
Aerospace Research and Development of US Air Force (BAM).

\end{multicols}

\end{document}